\documentclass[12pt]{iopart}
\usepackage{iopams}  

\usepackage{graphics}
\usepackage{graphicx}
\usepackage{bm}
\usepackage{amsfonts}
\usepackage{amssymb}
\usepackage{latexsym}

\begin{document}

\title{Sonoluminescence and quantum optical heating}
\author{Andreas Kurcz, Antonio Capolupo, and Almut Beige} 
\address{The School of Physics and Astronomy, University of Leeds, Leeds, LS2 9JT, United Kingdom}

\date{\today}

\begin{abstract}
Sonoluminescence is the intriguing phenomenon of strong light flashes from tiny bubbles in a liquid. The bubbles are driven by an ultrasonic wave and need to be filled with noble gas atoms. Approximating the emitted light by blackbody radiation indicates very high temperatures. Although
sonoluminescence has been studied extensively, the origin of the sudden energy
concentration within the bubble collapse phase is still controversial. It is
hence difficult to further increase the temperature inside the bubble for
applications like sonochemistry and table top fusion. 
Here we show that the strongly confined nobel gas atoms inside the bubble can be heated very rapidly by a weak but highly inhomogeneous electric field as it might occur naturally during rapid bubble deformations. An indirect proof of the proposed quantum optical heating mechanism
would be the detection of the non-thermal emission of photons in the optical regime {\em prior} to the light flash. Our model implies that it is possible to increase the temperature inside the bubble with the help of appropriately detuned laser fields.
\end{abstract}

\submitto{\NJP}
\maketitle

\section{Introduction}

In 1934, Frenzel and Schultes \cite{Frenzel} accidentally discovered a phenomenon that later became known as {\em multi-bubble sonoluminescence} \cite{Walton,Mcnamara}. In order to speed up the development process of pictures, they applied ultrasonic waves to a tank with a photographic
fluid. What they observed were tiny imploding bubbles emitting light at low intensity. Interest in this phenomenon increased again in 1989, when Gaitan {\em et al.} \cite{Gaitan} were able to produce the cavitation of a single bubble. Optimising the experimental setup, they created a stable bubble whose radius changed periodically in time. In each cycle, the bubble suddenly collapses and emits a sharp light pulse. In its new form, the phenomenon became known as {\em single-bubble sonoluminescence}. Later, the groups of Putterman \cite{Putterman,Camara,Barber1,Barber2,Barber3,Hiller,Vazquez,Vazquez2}, Suslick
\cite{Suslick,Suslick1,Suslick2,Suslick3,Suslick4,Flannigan2} and others \cite{Lohse1,Lohse2,Lee,Ciawi,Rae,Delmas,Burdin,Lauterborn} perfected these experiments. Luminescence of a cavitating bubble has even been induced by pulsed laser excitation \cite{Ohl}.

\begin{figure}
\begin{minipage}{\columnwidth}
{\includegraphics[]{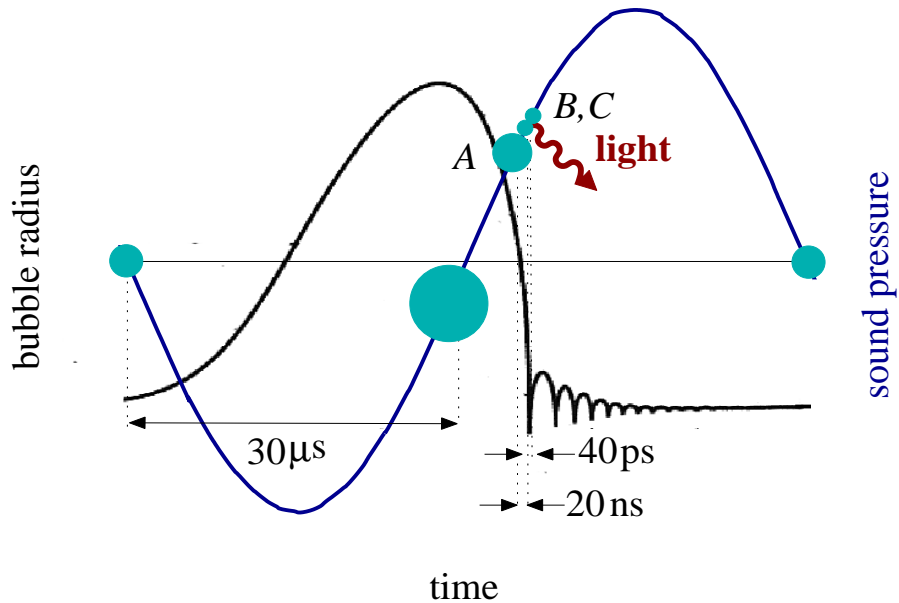}} 
\vspace*{-0.5cm} \caption{Time dependence of the driving sound pressure and of the bubble radius in a typical single-bubble sonoluminescence cycle. Point $A$ marks the beginning of the collapse phase in which the bubble approaches its minimum radius very rapidly and becomes thermally isolated from the liquid. In point $B$, the temperature within the bubble is significantly increased and a strong light flash occurs. Point $C$ denotes the beginning of the expansion phase in which the bubble oscillates around its equilibrium radius until it regains its stability.} \label{fig1}
\end{minipage}
\end{figure}

Fig.~\ref{fig1} shows a typical single-bubble sonoluminescence cycle and indicates the relevant time scales \cite{Brenner,Lohse}. The period of the applied sound wave is around $60\,\mu$s. The time dependence of the bubble radius is in good agreement with the laws of classical physics and can be described by the Rayleigh-Plesset equations \cite{Moss2}. For most of the cycle, the bubble radius increases isothermal. Each expansion phase is followed by a rapid collapse phase. Close to point $A$, i.e.~about $20 \,$ns before the minimum radius is reached, the accelerating bubble wall becomes so fast that the bubble becomes thermally isolated from the liquid. Close to its minimum radius of about $0.5 \, \mu$m, i.e.~between points $B$ and $C$, the bubble might be filled with up to $10^8$ noble gas atoms. At this point, a rapid increase of the energy density occurs which is accompanied by the sudden emission of light. In case of argon atoms, the light flash lasts for about $40\,$ps. Afterwards a re-expansion phase begins in which the bubble oscillates around its equilibrium radius until it regains its stability. 

Detailed measurements of the spectra of the picosecond light flash between point $B$ and $C$ in Fig.~\ref{fig1} have been made. Associating the continuum underlying these sonoluminescence spectra with blackbody or {\em Bremsstrahlung} radiation indicates temperatures of at least $10^3$--$10^4 \,$K inside the bubble \cite{Barber1,Barber3,Vazquez,Vazquez2,Suslick5,McNamara2}. It is even possible to observe light emission in the ultraviolet regime which hints at temperatures of about $10^6 \,$K \cite{Camara}. Noteworthy is the discovery of sharp emission lines in the optical regime \cite{Suslick3,Brenner,Suslick5}. These indicate the population of highly excited energy eigenstates of noble gas \cite{Suslick1,Suslick2,Suslick4,Flannigan2} and metal atoms \cite{Suslick3} which cannot be populated thermally. These excitations prove the formation of an opaque plasma core inside the bubble and have been observed in several single- and in multi-bubble sonoluminescence experiments \cite{Suslick4,Suslick5,Gordeychuk}. Recent experiments \cite{Suslick1,Suslick2,Suslick4,Flannigan2} already verified the existence of a dense plasma inside the bubble.

The light flash at the end of the bubble collapse phase has been attributed to surface blackbody radiation \cite{Barber3,Vazquez,Vazquez2,Hopkins}, neutral and ion Bremsstrahlung \cite{Wu,Moss3,Moss4,Xu}, collision-induced emission \cite{Frommhold,Frommhold2}, quantum vacuum radiation \cite{Schwinger,Eberlein}, and other thermal \cite{Hiller,Hilgenfeldt,Hilgenfeldt2} and non-thermal processes \cite{Bernstein,Willison,Garcia}. Other authors describe the plasma temperatures inside the bubble by a converging spherical shock wave \cite{Wu,Moss}. All of these models reproduce the observed single-bubble sonoluminescence spectra qualitatively. Nevertheless, the origin of the sudden energy concentration during the final stage of the bubble collapse phase in single-bubble sonoluminescence experiments remains a mystery \cite{Suslick5,Putterman2}. A valid theoretical model needs to include a mechanism for the formation of a plasma, as well as a mechanism which can increase the temperature of the plasma even further by at least one order of magnitude. This mechanism needs to be able to operate in a solid state-like environment and on the very small length scale given by the radius of the bubble.

In this paper we show that a relatively weak but highly inhomogeneous electric field can cause a coupling between the quantised motion and the electronic states of strongly confined particles. When combined with spontaneous emission, this coupling results in very high heating rates. The origin of this heating lies in the fact that the creation of a phonon is more likely than the annihilation of a phonon. This transfers the particles to higher and higher temperatures. We then point out that the electric field required for this process might occur naturally in sonoluminescence experiments during rapid bubble deformations. Moreover, the noble gas atoms inside the bubble form a van der Waals gas which reaches its van der Waals core when the bubble approaches its minimum radius \cite{Brenner,Suslick5}. They hence experience strong trapping potentials such that their motion becomes quantised. Furthermore we remark that noble gas atoms have a stable enough level configuration to undergo a strong heating process. The quantum optical heating of nobel gas atoms might therefore play a crucial role in single and multibubble sonoluminescence experiments. 

Our model does not contradict current models for the description of sonoluminescence experiments but might explain certain controversial aspects of this phenomenon. In addition to pointing out a previously unconsidered heating mechanism, it clarifies the role of the noble gas atoms. Indeed, sonoluminescence experiments require a relatively high concentration of them inside the bubble \cite{Brenner,McNamara2}. Moreover, our calculations show that the inhomogeneous electric field accompanying the heating process would result in a non-negligible population of excited atomic states and the non-thermal emission of light in the optical regime. The detection of a certain background radiation prior to the light flash during the bubble collapse phase would hence be a first indirect proof of our model. This radiation is expected to occur on a nanosecond time scale and should result in sharp emission lines at a typical noble gas transition. Its intensity should be proportional to the phonon frequency $\nu$ and to the number of emitting atoms but does not depend on their temperature. 

The model we use for the description of the proposed quantum optical heating mechanism is similar to the models typically used to describe laser sideband cooling in ion trap experiments \cite{Wineland,Eschner}. In these experiments a {\em red}-detuned laser field is applied. Like the above mentioned highly inhomogeneous electric field, it establishes a coupling between the quantised motion and the electronic states of trapped atoms. Based on this analogy, we conclude this paper with the proposal to enhance the energy concentration in sonoluminescence experiments with the help of appropriately {\em blue}-detuned laser fields. The observation of a certain dependence of the sonoluminescence phenomenon on the frequency (and lesser on the intensity) of the applied laser field would support our thesis of quantum optical heating. Increasing the temperatures inside the bubble via laser driving could  assist sonochemistry in the synthesis of a wide range of nanostructured materials and in the preparation of biomaterials \cite{Sonochem1,Sonochem2,Sonochem3} and might even help to facilitate nuclear fusion \cite{Barber4,Suslick7}.

There are seven sections in this paper. In Section \ref{perspective}, we explain the basic idea behind the considered quantum optical heating mechanism. A theoretical description of strongly confined atoms inside a highly inhomogeneous electric field is given in Section \ref{dynamics}. In Section \ref{hotter} we solve the time evolution of a single atom which is typical for all the atoms inside the bubble. In Section \ref{B} we discuss the emission of a certain background radiation in the optical regime which accompanies the heating process  prior to the light flash. Section \ref{poss} concludes our discussion by pointing out a mechanism to further increase the temperature inside the bubble. We finally summarise our results in Section \ref{conclusions}.

\section{A quantum optical heating mechanism} \label{perspective}

In this paper, we describe a quantum optical heating mechanism which occurs when a weak but highly inhomogeneous electric field interacts with strongly confined noble gas atoms. Such a field might occur naturally during the sudden contractions of the bubble radius in single-bubble sonoluminescence experiments and during rapid bubble movements and deformations in multi-bubble sonoluminescence experiments. Polar molecules of the liquid dissolve and ionized species might be trapped inside the bubble \cite{Brenner,Suslick5,Didenko}, thereby leading to a non-negligible inhomogeneous charge distribution. This charge distribution is further enhanced by the formation of a plasma which might create a significant excess of charge density \cite{Suslick1,Suslick2,Suslick4,Flannigan2,Suslick5,Gordeychuk}. Although, the electric field gradient inside the bubble might be negligible for most of the time, it might become very high on the relevant $\mu$m length scale which is given by the bubble radius, when the bubble approaches its light emission stage.

When the bubble reaches its minimum radius (cf.~point $B$ in Fig.~\ref{fig1} in single-bubble sonoluminescence experiments), the mean distance between the noble gas atoms becomes so small that interactions between them can be described by a Lennard-Jones potential. The atoms experience an equilibrium between repulsive interatomic forces due to overlapping orbitals (also referred to as Pauli repulsion) and attractive forces. To model the resulting strong confinement of the atoms inside the bubble, we place each of them into a trapping potential. This allows us to quantise the atom motion during the collapse phase when the bubble reaches its maximum compression. For simplicity we restrict ourselves to the assumption of only one phonon mode per atom and denote its frequency by $\nu$. We expect that the nobel gas atoms have a stable enough level configuration to undergo strong heating processes.

In the next section, we shall see that the gradient of the electric field inside the bubble establishes a coupling between the quantised motion and the electronic states of each noble gas atom. To model this we denote the ground state of each noble gas atom by $|0 \rangle$ while $|1 \rangle$ is an excited electronic state. The interaction Hamiltonian contains terms that correspond to the excitation and to the de-excitation of atoms accompanied by the creation and the annihilation of a phonon. Crucial for changing the temperature of the atoms is the presence of a large spontaneous decay rate of the excited state $|1\rangle$ which keeps the atoms predominantly in their ground state. Although the described  processes are highly non-resonant, we show below that they can result in a significant change of the mean phonon number per atom. Within a few nanoseconds, the temperature inside the bubble can increase by many orders of magnitude. 

\begin{figure}
\begin{minipage}{\columnwidth}
{\includegraphics[]{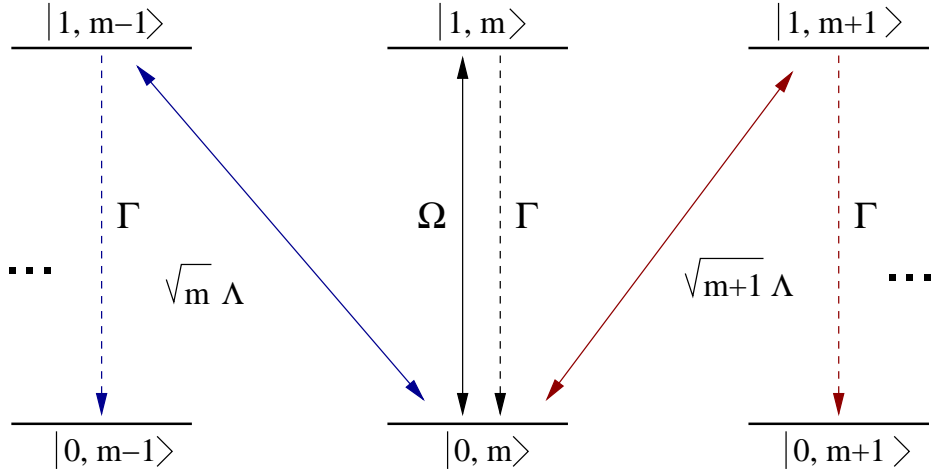}}
\caption{Level configuration of a single atom-phonon system indicating the immediately relevant transitions, if the atom is initially in its ground state $|0 \rangle$ and possesses exactly $m$ phonons. As in Eq.~(\ref{Hint}), $\Omega$ denotes the coupling constant for phonon conserving transitions to the excited atomic state $|1 \rangle$, while $\Lambda $ is due to the electric field gradient inside the bubble and establishes a coupling between the electronic and the motional states of the atom. Moreover, $\Gamma$ is the spontaneous photon decay rate of level 1.} \label{fig2}
\end{minipage}
\end{figure}

Before going into a more detailed analysis, let us describe the proposed heating mechanism in a more intuitive way. Suppose an atom is initially in its ground state and possesses exactly $m$ phonons. We denote this state by $|0,m \rangle$. Fig.~\ref{fig2} shows the immediately relevant transitions for this state which we derive explicitly in the next section (cf.~Eqs.~(\ref{Hint}) and (\ref{master2})). Notice that phonons are bosons with their annihilation and creation operator $b$ and $b^\dagger$ given by
\begin{eqnarray} \label{bs} 
b = \sum_{m=1}^\infty \sqrt{m} \, |m - 1 \rangle \langle m| ~~& {\rm and} &~~
b^\dagger = \sum_{m=0}^\infty \sqrt{m+1} \, |m + 1 \rangle \langle m| 
\end{eqnarray}
with $[b,b^\dagger ] = 1$. Consequently, a transition from the state $|0,m \rangle$ into $|1, m+1 \rangle$ occurs with a rate proportional to $\sqrt{m+1}$, while the rate for a transition from $|0,m \rangle$ into $|1, m-1 \rangle$ scales as $\sqrt{m}$. When the spontaneous decay rate of the atom is relatively large, such a transition is most likely followed by an irreversible and predominantly non-radiative transition back into the ground state of the atom. It transfers the atom either into its initial state $|0,m \rangle$, into $|0, m-1 \rangle$, or into $|0, m+1 \rangle$. Since the final population in the state with $m+1$ phonons is higher than the population in the state with $m-1$ phonons, the net effect of the described excitation and de-excitation process is an increase of the mean phonon number per atom, i.e.~heating.

\section{Theoretical model} \label{dynamics}

Let us now have a closer look at the mechanism which creates the coupling between the quantised motion and the electronic states of the nobel gas atom shown in Fig.~\ref{fig2}. To do so we assume the presence of a weak but highly inhomogeneous electric field and derive effective rate equations for the time evolution of a {\em single} atom-phonon system which is typical for the many atoms inside the bubble. For simplicity, we approximate each atom by a two-level system.

\subsection{The interaction Hamiltonian}

Suppose $\sigma^+ \equiv |1 \rangle \langle 0|$, $\sigma^- \equiv |0 \rangle \langle 1|$ and ${\bf r}$ is the position of a single atom inside the bubble. Within the usual dipole approximation, the Hamiltonian of the atom is given by the dipole interaction
\begin{eqnarray} \label{Hs}
H _{\rm int} &=& e\, {\bf D} \cdot {\bf E} ({\bf r})  
\end{eqnarray}
with $e$ being the charge of a single electron, the (real) atomic dipole moment 
\begin{eqnarray}
{\bf D} &=&  {\bf D}_{01} \, \sigma^{-} + {\rm H.c.} \, ,
\end{eqnarray}
and the electric field ${\bf E} ({\bf r})$. For simplicity, we assume that all field components point in the same direction of a unit vector $\hat {\bf k}$, which allows us to write ${\bf E} ({\bf r})$ as 
\begin{eqnarray}
{\bf E} ({\bf r}) &=& \sum_k {\bf E}_k \, {\rm e}^{{\rm i} k \hat {\bf k} \cdot {\bf r}} + {\rm c.c.} 
\end{eqnarray}  
with amplitudes ${\bf E}_k$ and wave vectors ${\bf k}= k \hat {\bf k}$. Moreover, we consider the motion of the atom as quantised. For simplicity, we describe its motion in the $\hat {\bf k}$-direction by only a single phonon mode with frequency $\nu $ and the annihilation operator $b$ in Eq.~(\ref{bs}). Replacing the position operator ${\bf r} - {\bf R}$ accordingly, we find \cite{LambDicke}
\begin{eqnarray} \label{before}
{\bf k} \cdot ({\bf r} - {\bf R}) &=& k \Delta x \big( b  + b^\dagger \big) \, ,
\end{eqnarray}
where ${\bf R} $ is the current equilibrium position of the atom, $M$ is its mass, and
\begin{eqnarray} \label{shrink3}
\Delta x &=& (\hbar / 2 M \nu)^{1/2} 
\end{eqnarray}
would be the width of its ground state wave function when replacing the square-well like potential seen by each atom by a harmonic one. An accurate description of the motion of the atom would involve a continuous range of phonon frequencies. Here $\nu $ is the dominating frequency of the populated phonon modes. This approximation is similar to the approximation made when describing the light emission from a blackbody by a single frequency in order to attribute a temperature $T$.

Eq.~(\ref{shrink3}) can be used to get a {\em very rough} estimate of the typical phonon frequency $\nu$ in sonoluminescence experiments. Suppose, the bubble has a radius of $500 \,$nm and is filled with about $10^7$ argon atoms. Although this might vary by at least one order of magnitude, we assume that the volume of each atom is given by $\Delta x^3$. Comparing this volume with $V/N$, where $V$ is the volume of the bubble, we obtain a typical $\Delta x$ of about $4 \,$nm. Using Eq.~(\ref{shrink3}) and taking as an example the atomic mass of argon into account, we find that the phonon frequency $\nu$ is of the order of $50 \,$MHz. This frequency is roughly of the same order of magnitude (or higher) as the phonon frequency in typical ion trap experiments \cite{Wineland,Eschner}.

Since the atom is well localized within the wavelength of its trapping potential, i.e.~$k \Delta x \ll 1$, we can now apply the Lamb-Dicke approximation which is routinely used in the theoretical modelling of ion trap experiments. It allows us to assume \cite{LambDicke}
\begin{eqnarray}
\exp( {\rm i} k \hat {\bf k} \cdot ({\bf r}-{\bf R})) &=& 1 + {\rm i} k \Delta x \big(b + b^\dagger \big) \, .
\end{eqnarray}
Substituting this into Eq.~(\ref{Hs}), we obtain the interaction Hamiltonian 
\begin{eqnarray} \label{Hint}
H _{\rm int} &=& \hbar \Omega \, ( \sigma ^- + \sigma^+) + \hbar \Lambda  \, (b  + b ^\dagger) ( \sigma ^- + \sigma^+)
\end{eqnarray}
with the (real and positive) coupling constants 
\begin{eqnarray} \label{shrink}
\Omega &\equiv & (2 e / \hbar) \sum _k {\bf D}_{01} \cdot {\rm{Re}} \left ( {\bf E}_k {\rm e}^{{\rm{i}} k \hat {\bf k} \cdot \bf{R}} \right ) \, , \nonumber \\
\Lambda &\equiv & - (2 e \Delta x / \hbar) \sum _k k  \, {\bf D}_{01} \cdot {\rm{Im}} \left ( {\bf E}_k {\rm e}^{{\rm{i}} k \hat {\bf k} \cdot {\bf R}} \right )\, .
\end{eqnarray}
This Hamiltonian is essentially a Jaynes-Cummings Hamiltonian \cite{Knight}. A closer look at Eq.~(\ref{shrink}) shows that $\Lambda$ is proportional to the gradient of $\Omega$ in the direction of the quantised motion of the atom, i.e.
\begin{eqnarray} \label{shrink2}
\Lambda &=& \Delta x \, \hat{\bf k} \cdot \nabla \Omega ({\bf R}) \, . 
\end{eqnarray}
A strong atom-phonon coupling therefore does not necessarily require the presence of a strong electric field. It only requires a highly inhomogeneous electric field inside the bubble. 

\subsection{The master equation}

Dissipation in the form of spontaneous photon emission from the excited electronic state $|1 \rangle$ is as usual taken into account by the master equation \cite{Knight}
\begin{eqnarray} \label{master2}
\dot \rho &=& - \frac{{\rm i}}{\hbar} \big[ H _{\rm int} + \hbar \omega_0 \, \sigma ^+ \sigma ^- + \hbar \nu  \, b ^\dagger b  \, ,  \rho \big] + \Gamma \big[ \, \sigma ^- \, \rho \, \sigma ^+ - {\textstyle \frac{1}{2}} \sigma ^+ \sigma ^- \, \rho \nonumber \\
&&  -  {\textstyle \frac{1}{2}} \, \rho \, \sigma ^+ \sigma ^- \, \big] \, .
\end{eqnarray}
Here $\hbar \omega_0$ is the energy of the excited state $|1 \rangle$ with the spontaneous decay rate $\Gamma$. Instead of integrating the master equation, we now use it to obtain a closed set of effective rate equations which can be solved more easily. Notice that some of the major quantities in these equations are coherences and not populations. 

\subsection{Effective rate equations}

In this paper we are especially interested in the mean phonon number per atom and the population in the excited state $|1 \rangle$. These are given by
\begin{eqnarray} \label{coh}
m \equiv \langle b ^\dagger  b  \rangle ~~& {\rm and} &~~ X_3 \equiv \langle \sigma^+ \sigma^- - \sigma^- \sigma^+ \rangle \, .
\end{eqnarray}
As we shall see below, a crucial role in their time evolution is played by the expectation values 
\begin{eqnarray} \label{cos}
&& X_1 \equiv \langle  \sigma^+ + \sigma^- \rangle \, , ~~
X_2 \equiv {\rm i} \langle \sigma^- - \sigma^+ \rangle \, , \nonumber \\
&& Y_1 \equiv \langle b + b^\dagger \rangle \, , ~~ 
Y_2 \equiv {\rm i} \langle b - b^\dagger \rangle \, , ~~ \nonumber \\
&& Y_3 \equiv \langle b^2 + b^{\dagger \, 2} \rangle \, , ~~
Y_4 \equiv {\rm i} \langle b^2 - b^{\dagger \, 2} \rangle 
\end{eqnarray} 
and the atom-phonon coherences
\begin{eqnarray} \label{cos2}
&& Z_1 \equiv \langle (\sigma^+ + \sigma^-) ( b + b^\dagger ) \rangle \, , ~~ 
Z_2 \equiv {\rm i} \langle (\sigma^- - \sigma^+) ( b + b^\dagger ) \rangle \, , ~~ \nonumber \\
&& Z_3 \equiv {\rm i} \langle (\sigma^+ + \sigma^-) ( b - b^\dagger ) \rangle \, , ~~
Z_4 \equiv - \langle (\sigma^- - \sigma^+) ( b - b^\dagger ) \rangle \, .
\end{eqnarray} 
To obtain a closed and relatively simple system of rate equations, we assume in the following that
\begin{eqnarray} \label{omega0}
\omega_0 \, \gg \, \nu, \, \Gamma, \, \Omega, \, \Lambda ~~& {\rm and} &~~ m \gg 1 \, .
\end{eqnarray}
Moreover, we approximate the expectation value of operators of the form
$\langle B \sigma _3 \rangle$ by $\langle B \, \rangle \langle \sigma _{3}
\rangle$, as it applies when the expectation value of $B$ is about the same
for an atom in $|0 \rangle$ and for an atom in $|1 \rangle$. Doing so, we find that the time derivatives of the variables in Eqs.~(\ref{coh})--(\ref{cos2}) are given by  
\begin{eqnarray} \label{fullset}
&& \dot m = \Lambda  Z_3 \, , ~~  
\dot X_3 = 2 (\Omega X_2 + \Lambda Z_2 ) - \Gamma \left ( X_3 + 1 \right ) \, , ~~ \nonumber \\
&& \dot Y_1 = - \nu Y_2 \, , ~~
\dot Y_2 = 2 \Lambda X_1 + \nu Y_1 \, , ~~ \nonumber \\
&& \dot Y_3 = - 2 (\nu Y_4 + \Lambda Z_3) \, , ~~  
\dot Y_4 = 2 (\nu Y_3 + \Lambda Z_1) \, ,
\end{eqnarray}
while
\begin{eqnarray} \label{fullset2}
&& \dot X_1 = - \omega_0 X_2 \, , ~~ 
\dot X_2 = - 2 (\Omega + \Lambda Y_1) X_3 + \omega_0 X_1 \, , ~~ \nonumber \\
&& \dot Z_1 = - \omega_0 Z_2 \, , ~~
\dot Z_2 = - 2 (\Omega Y_1 + \Lambda Y_3 +2 \Lambda m) X_3 + \omega_0 Z_1 \, , ~~ \nonumber \\
&& \dot Z_3 =  2 \Lambda - \omega_0 Z_4 \, , ~~
\dot Z_4 = - 2 (\Omega Y_2 + \Lambda Y_4 ) X_3 + \omega_0 Z_3 \, .
\end{eqnarray}
These rate equations hold up to first order in $1/\omega_0$. 

Taking into account that $\omega_0$ is much larger than all other system parameters (cf.~Eq.~(\ref{omega0})), we find that the rate equations in Eq.~(\ref{fullset2}) can be eliminated adiabatically from the full set of rate equations. Doing so, $X_1$, $X_2$, and the $Z$ coherences become
\begin{eqnarray} \label{fullset3}
&& X_1 = {2 (\Omega + \Lambda Y_1) X_3 \over \omega_0} \, , ~~ X_2 = 0 \, , ~~ \nonumber \\
&& Z_1 = {2 (\Omega Y_1 + \Lambda Y_3 +2 \Lambda m) X_3 \over \omega_0} \, , ~~ Z_2 = 0 \, , \nonumber \\
&& Z_3 =  {2 (\Omega Y_2 + \Lambda Y_4 ) X_3 \over \omega_0} \, , ~~ Z_4 = {2 \Lambda \over \omega_0} 
\end{eqnarray}
in first order in $1/\omega_0$. These quasi-stationary state values are reached on the very fast time scale given by $1/\omega_0$ which is typically of the order of femtoseconds. Substituting them into Eq.~(\ref{fullset}) we finally obtain the effective rate equations 
\begin{eqnarray} \label{R1}
&& \dot X_3 = - \Gamma \left ( X_3 + 1 \right ) \, , \nonumber \\
&&  \dot Y_1 = - \nu Y_2 \, , ~~ \dot Y_2 = {4 \Lambda (\Omega + \Lambda Y_1) X_3 \over \omega_0} + \nu Y_1 \, .
\end{eqnarray}
Moreover we have
\begin{eqnarray} \label{R2}
\dot m &=& {2 \Lambda (\Omega Y_2 + \Lambda Y_4 ) X_3 \over \omega_0} \, , \nonumber \\
\dot Y_3 &=& - {4 \Lambda (\Omega Y_2 + \Lambda Y_4 ) X_3 \over \omega_0} - 2 \nu Y_4  \, , \nonumber \\
\dot Y_4 &=& {4 \Lambda (\Omega Y_1 + \Lambda Y_3 +2 \Lambda m) X_3 \over \omega_0} + 2 \nu Y_3 \, .
\end{eqnarray}
Notice that the three differential equations (\ref{R1}) decouple from the rest and can be solved separately.

\section{The dynamics of the system} \label{hotter}

In the following, we consider two cases. First we discuss the case of time-independent system parameters in the strong atom-phonon coupling regime with
\begin{eqnarray} \label{par} 
\Lambda \, \gg \, \Omega  ~~& {\rm and} &~~ 4 \Lambda^2 \, > \, \nu \omega_0 
\end{eqnarray}
and show that the effective rate equations (\ref{R1}) and (\ref{R2}) can
indeed result in a very strong and approximately exponential heating
process. We also identify the most important quantities in the time evolution
of the system. In actual sonoluminescence experiments, the system parameters can change rapidly in time, especially during the final stage of the bubble collapse phase. As we shall see below, rapidly changing parameters might enhance the heating of the atom inside the bubble
even further. 

\subsection{Time-independent system parameters} \label{initial}

\begin{figure}
\begin{minipage}{\columnwidth}
{\includegraphics[]{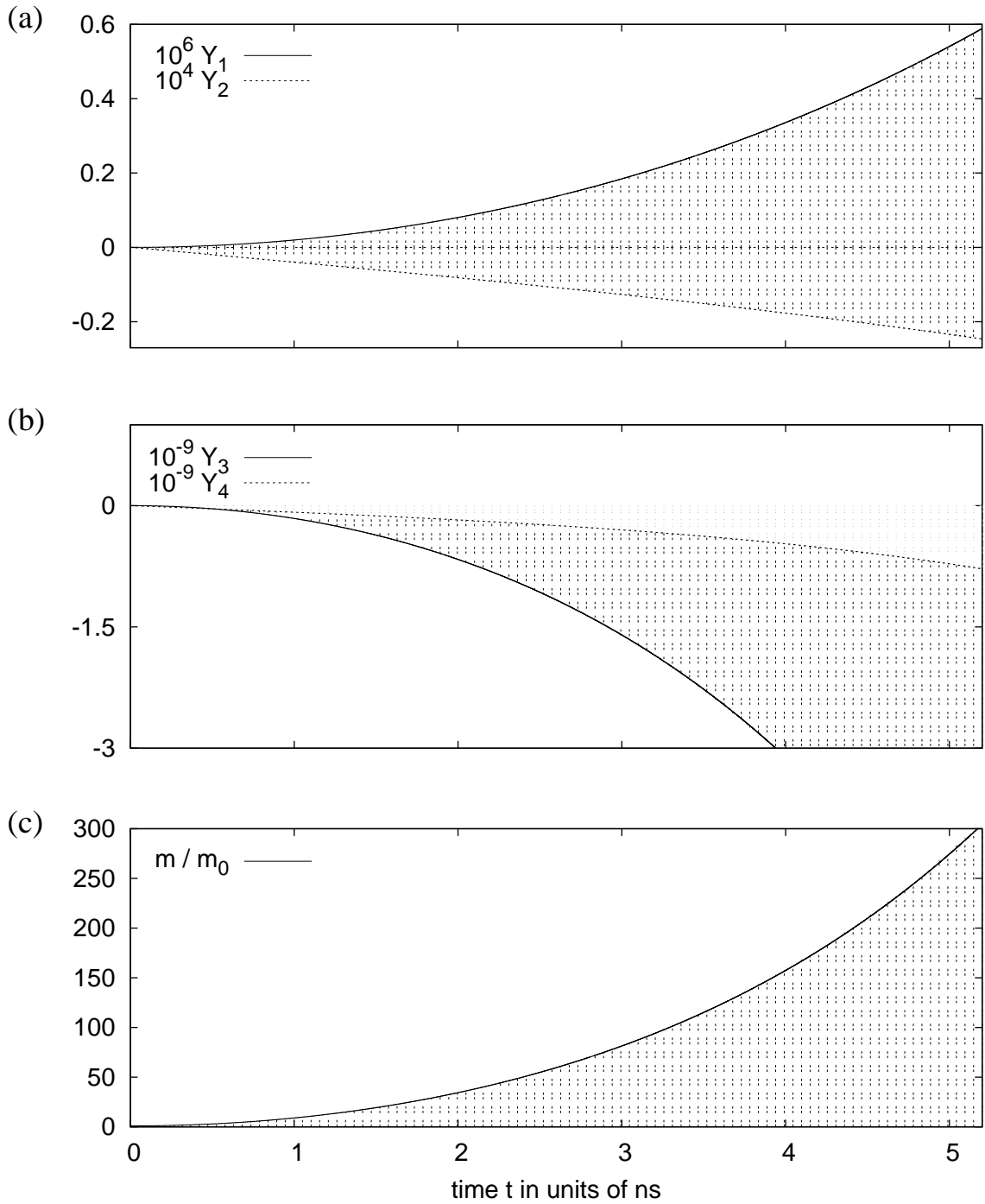}}
\vspace*{-0.5cm} \caption{The $Y$ coherences and the mean phonon number $m$ per atom as a function of
  time for $\nu = 10\,$MHz while $\Omega = 10^{6} \, $Hz, $\Lambda = 10^{12}
  \, $Hz, $\Gamma = 10^{13} \, $Hz \cite{footnote}, and $\omega_0 = 10^{15} \,
  $Hz obtained from a numerical solution of the full rate equations
  (\ref{fullset}) and (\ref{fullset2}). Good agreement is found with the
analytical solutions in Eqs.~(\ref{X32}), (\ref{fullsolution}), and (\ref{fullsolution2}) which are represented by the shaded areas.} \label{fig3}
\end{minipage}
\end{figure}

\begin{figure}
\begin{minipage}{\columnwidth}
{\includegraphics[]{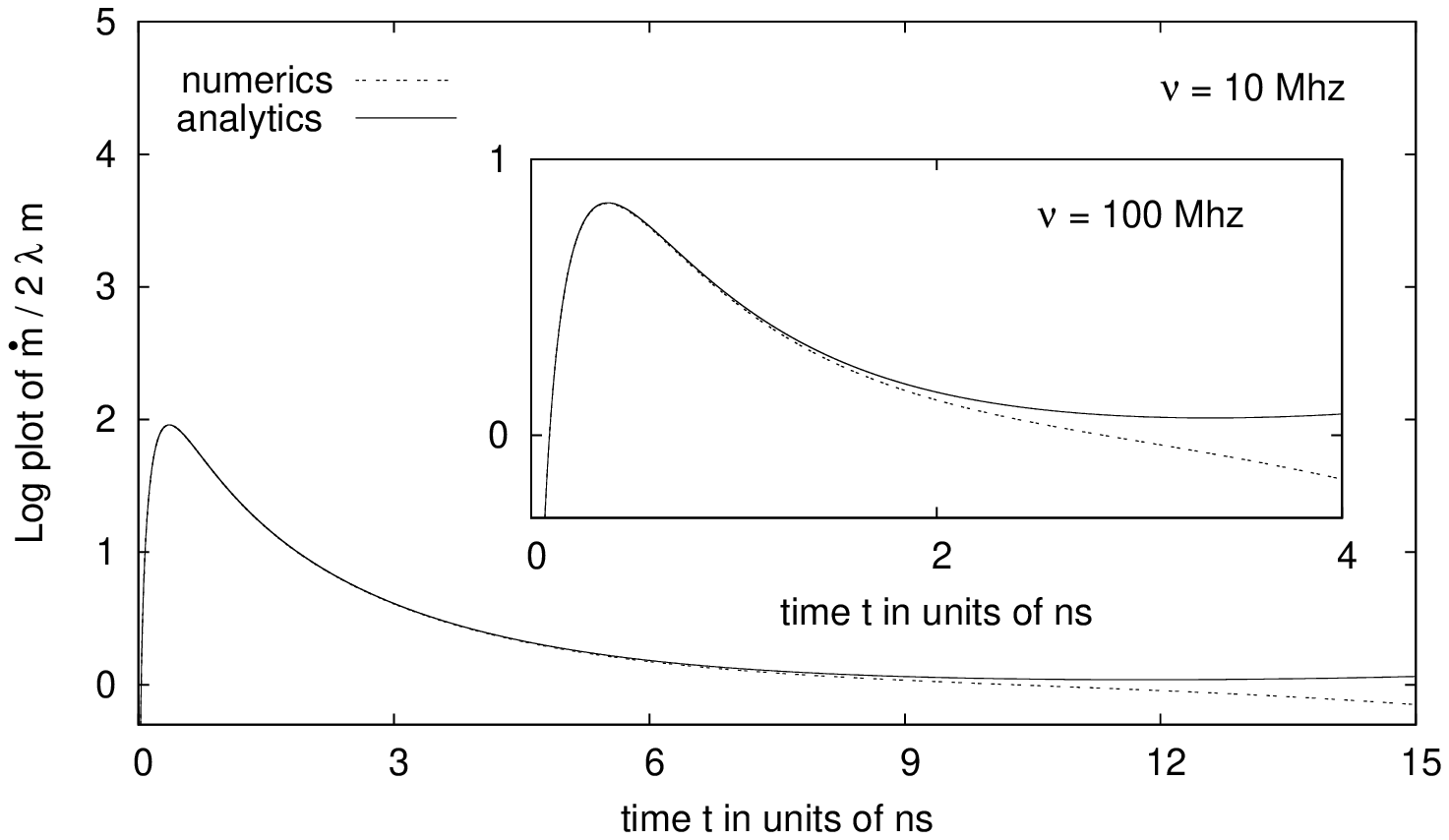}} 
\vspace*{-0.5cm} \caption{Logarithmic plot of the heating rate $\dot m/m$ as a function of time in units of $2 \lambda$ for $\nu = 10\,$MHz and $\nu = 100\,$MHz and for $\omega_0$, $\Omega$, $\Lambda$, and $\Gamma$ as in Fig.~\ref{fig3}. Again, good agreement is found between the numerical solution of the full rate equations which can be found in Eqs.~(\ref{fullset}) and (\ref{fullset2}) and the analytical solution (cf.~Eqs.~(\ref{fullsolution})--(\ref{mdot})).} \label{fig4}
\end{minipage}
\end{figure} 

In the beginning of each sonoluminescence cycle, the atoms experience
neither a strong trapping potential nor an inhomogeneous electric field. We
can therefore assume that the coherences defined in Eq.~(\ref{cos}) are
initially zero and that the atom is in its ground state. Due to having previously been in a hot environment, the mean phonon number per atom, $m_0$, is in general already relatively high when the trap is formed during the bubble collapse phase. 
Solving Eq.~(\ref{R1}) for these initial conditions and the strong coupling regime assumed in Eq.~(\ref{par}) yields 
\begin{eqnarray} \label{X3}
X_3 (t) &=& -1
\end{eqnarray}
up to first order in $1/\omega_0$. Moreover, we find in good agreement with a numerical solution of the full rate equations (cf.~Fig.~\ref{fig3}(a)) that
\begin{eqnarray} \label{X32}
Y_1 (t) &=& \frac{4 \nu \Omega \Lambda}{\lambda^2 \omega_0} \, \big[ \cosh ( \lambda t ) -1 \big] \, , ~~  \nonumber \\
Y_2 (t) &=& -  \frac{4 \Omega \Lambda}{\lambda \omega_0} \, \sinh ( \lambda t )  
\end{eqnarray}
with the (real) frequency $\lambda$ defined as
\begin{eqnarray} \label{lambda}
\lambda &\equiv & \nu \left( {4 \Lambda^2 \over \nu \omega _0} - 1 \right)^{1/2} \, .
\end{eqnarray}
Eq.~(\ref{X3}) shows that the atom remains predominantly in its ground state. Its role is to act as a catalyser for the heating process described in Section \ref{perspective}. For times $t$ of the order of $1/\lambda$, $Y_1$ and $Y_2$ are only of the order of $1/\omega_0$. We can hence safely assume that 
\begin{eqnarray}
\Omega Y_2 \ll \Lambda Y_4 ~~& {\rm and} &~~ \Omega \ll 2 \Lambda m 
\end{eqnarray}
on a timescale of a few nanoseconds. Taking this and Eq.~(\ref{X3}) into account, we find that the expressions for $Z_1$ and $Z_3$ in Eq.~(\ref{fullset3}) simplify to 
\begin{eqnarray}
Z_1 = - 2 \Lambda (2 m + Y_3)/\omega_0 ~~& {\rm and} &~~ Z_3 =  - 2 \Lambda Y_4/\omega_0
\end{eqnarray}
in first order in $1/\omega_0$. The remaining variables $m$, $Y_3$, and $Y_4$ in Eq.~(\ref{fullset}) hence evolve according to  
the differential equation
\begin{eqnarray}\label{set}
\dot {\bf v} &=& M {\bf v}
\end{eqnarray}
with ${\bf v} \equiv (m ,Y_3, Y_4)^{\rm T}$ and 
\begin{eqnarray}\label{set2}
M \equiv - {2 \over \omega_0} \left(\begin{array}{ccc} 0 & 0 & \Lambda^2 \\ 0 & 0 & -(2 \Lambda^2 - \nu \omega_0) \\ 4 \Lambda^2 & 2 \Lambda^2 - \nu \omega_0 & 0 \end{array} \right) \, .  ~~
\end{eqnarray}
Solving this equation for $m(0)=m_0$ and $Y_3(0)=Y_4(0)=0$, we find 
\begin{eqnarray} \label{fullsolution}
Y_3(t) &=& - {8 \Lambda^2 (2 \Lambda^2 - \nu \omega_0) \over \lambda^2 \omega_0^2} \, m_0 \, \sinh^2 ( \lambda t ) \, , \nonumber \\
Y_4(t) &=& - {4 \Lambda^2 \over \lambda \omega_0} \, m_0 \, \sinh ( 2 \lambda t ) \, ,
\end{eqnarray}
which is in good agreement with the numerical solution of the full rate equations in Figs.~\ref{fig3}(b) and \ref{fig3}(c). The mean number of phonons per atom evolves according to
\begin{eqnarray} \label{fullsolution2}
m(t) &=& m_0 + {8 \Lambda^4 \over \lambda^2 \omega _0^2} \, m_0 \, \sinh^2 ( \lambda t ) \, .
\end{eqnarray}
Fig.~\ref{fig3}(c) shows that this equation indeed describes an approximately exponential heating process. However, as shown in Fig.~\ref{fig4}, the relative heating rate $\dot m/m$ is not constant in time but has a relatively strong time dependence itself. At $t=0$, we have $\dot m / m = 0$. It then increases rapidly and assumes values well above $2 \lambda$  after a very short time.  From Eqs.~(\ref{set}) and (\ref{set2}) we see that
\begin{eqnarray} \label{mdot}
{\dot m \over m} &=& - {2 \Lambda^2 \over \omega_0} \, {Y_4 \over m} \, .
\end{eqnarray}
For times $t$ of the order of  $1/\lambda$, the relative heating rate $\dot m/m$ hence approaches $2 \lambda$, since
the expressions $2 \sinh^2 (\lambda t)$ and $\sinh (2 \lambda t)$ in Eq.~(\ref{fullsolution}) are of about the same size in this case.

\subsection{Time-dependent system parameters}

Up to now we considered only time-independent system parameters. However, in
actual sonoluminescence experiments, system parameters like the phonon
frequency $\nu$ and the atom-phonon coupling constant
$\Lambda$ might change very rapidly in time. The reason for this is that
$\Lambda $ originates from a highly inhomogeneous electric field which depends, for example, strongly on the bubble radius. The phonon frequency $\nu$ also relates to the radius of the bubble since it depends on
the trapping potential experienced by each atom. To take this into
account, we now analyse the effect of time-dependent parameters $\nu(t)$ and
$\Lambda(t)$. The concrete size of $\Gamma$ and $\Omega$ is less crucial for the time evolution of the mean phonon number $m$, as we can see from the absence of these two parameters in Eq.~(\ref{set2}).

Let us assume that $\nu$ and $\Lambda$ change on a time scale $t$ which is long compared to the very short time scale given by $1/\omega _0$. Then the above adiabatic elimination of the differential equations (\ref{fullset2}) remains valid and the dynamics of the atom-phonon system is given by Eqs.~(\ref{R1}) and (\ref{R2}). Moreover, we have seen in the previous section that $Y_1$ and  $Y_2$ effectively do not contribute to the time evolution of $m$, $Y_3$, and $Y_4$. Their dynamics is hence to a very good approximation given by Eq.~(\ref{set}). In the following, we solve this differential equation for rapidly changing system parameters $\nu$ and $\Lambda$ for different situations analytically and numerically. 

To do so, we first notice that the eigenvalues $\lambda_i$ of $M$ are given by
\begin{eqnarray}
\lambda_1 =0 ~~& {\rm and} &~~ \lambda_{2,3} = \pm 2 \lambda 
\end{eqnarray}
with $\lambda$ as in Eq.~(\ref{lambda}). The corresponding eigenvectors ${\bf u}_i$ of $M$ are given by 
\begin{eqnarray}
{\bf u}_1 &=& {1 \over {\sqrt{2} \sqrt{\nu ^4 + \lambda ^4}}} \left(
  \begin{array}{c} \nu ^2 - \lambda ^2 \\ 2
    ( \nu ^2+ \lambda ^2 ) \\ 0 \end{array} \right) \, , \nonumber \\
{\bf u}_{2,3} &=& {1 \over {\sqrt{5} (\nu ^2 + \lambda ^2)}} \left( \begin{array}{c} \mp
    (\nu ^2 + \lambda ^2) \\ \pm 2 ( \lambda ^2 - \nu ^2 ) \\ 4 \nu \lambda
  \end{array} \right) \, .
\end{eqnarray}
These eigenvectors are in general not pairwise orthogonal. Our calculation below therefore involves also their dual vectors ${\bf u}^i$ which equal
\begin{eqnarray}
{\bf u}^1 &=& {\sqrt{\nu ^4 + \lambda ^4} \over {4 \sqrt{2} (\lambda \nu )^2}} \left(
  \begin{array}{c} 2 (\lambda ^2 - \nu ^2)\\ \nu^2 + \lambda^2 \\ 0 \end{array} \right) \, , \nonumber \\
{\bf u}^{2,3} &=& {\sqrt{5} \over 16} {\nu^2 + \lambda ^2 \over {(\lambda \nu) ^2 }} \left( \begin{array}{c}
    \mp 2 (\nu^2 + \lambda ^2 )\\ \mp (\lambda ^2 - \nu^2) \\ 2 \nu \lambda \end{array} \right) 
\end{eqnarray}
with ${\bf u}^{j} \cdot {\bf u}_{i}= \delta _{ij}$. This means, the dual vector ${\bf u}^j$ is orthogonal to the two vectors ${\bf u}_i$ with $i \neq j$ and its overlap with ${\bf u}_j$ is unity.

In the following, we write the vector ${\bf v} = (m,Y_3,Y_4)^{\rm T}$ in Eq.~(\ref{set}) as a superposition of the eigenvectors of $M$. Denoting the respective coefficients by $c_i$, i.e. 
\begin{eqnarray} \label{shrink6}
{\bf v} &=& \sum_{i=1}^3 c_i \, {\bf u}_i \, , 
\end{eqnarray}
using the fact that $M \, {\bf u}_i = \lambda_i \, {\bf u}_i$, and taking the time dependence of the eigenvectors of $M$ into account, we can then show that 
\begin{eqnarray} \label{transformer}
\dot {\bf v} \, = \, \sum_{i=1}^3  \dot c _i \, {\bf u}_i + c_i \, \dot {\bf u}_i \, = \, \sum_{i=1}^3 c _i \lambda _i \, {\bf u_i} \, .
\end{eqnarray}
Multiplying this equation with the ${\bf u}^j$ finally yields 
\begin{eqnarray} \label{dotcj}
\dot c_j & = & \lambda_j c_j - \sum_{i=1}^3 c _i \, {\bf u}^j \cdot \dot {\bf u}_i \, .
\end{eqnarray}
This shows that the coefficients $c_i$ evolve according to 
\begin{eqnarray} \label{dotcj2}
\dot c_1 & = & \left[ { \lambda^4 - \nu^4
    \over \lambda^4 + \nu^4} \, c_1 - \sqrt{2 \over 5} {\sqrt{\nu^4 + \lambda ^4} \over \nu^2 + \lambda^2} \, (c_2 - c_3) \right] \, {{\rm d} \over {\rm d}t} \ln \left ( {\lambda \over \nu} \right ) \, , \nonumber \\
\dot c_2 & = & \left[ { \lambda^2 - \nu^2
    \over \lambda^2 + \nu^2} \, c_2 - \sqrt{5 \over 8} {\nu^2 + \lambda ^2 \over \sqrt{\nu^4 +
      \lambda^4}} \, c_1 \right] \, {{\rm d} \over {\rm d}t} \ln \left ( {\lambda \over \nu} \right ) + 2 \lambda \, c_2 \, , \nonumber  \\
\dot c_3 & = & \left[ { \lambda^2 - \nu^2
    \over \lambda^2 + \nu^2} \, c_3 + \sqrt{5 \over 8} {\nu^2 + \lambda ^2 \over \sqrt{\nu^4 +
      \lambda^4}} \, c_1  \right] \, {{\rm d} \over {\rm d}t} \ln \left ({\lambda \over \nu} \right ) - 2 \lambda \, c_3 \, .
\end{eqnarray}
These differential equations take the effect of the rapidly changing parameters $\lambda$ (which depends on $\nu$ and $\Lambda$) and $\nu$ into account and can be used to analyse the dynamics of ${\bf v}$ (i.e.~$m$, $Y_3$, and $Y_4$) for times $t$ of the order of $1/\lambda$. Unfortunately, their analytical solution is in general not straightforward.

To illustrate that rapidly changing system parameters generally do not suppress the above described heating process, we now consider the following scenario as an example. Let us assume that $\lambda$ and $\nu$ change such that their ratio $\eta$,
\begin{eqnarray} \label{shrink8}
\eta &\equiv & \lambda / \nu \, ,
\end{eqnarray}
remains about the same. Using Eqs.~(\ref{shrink3}), (\ref{shrink2}) and (\ref{lambda}), we see that $\eta$ equals 
\begin{eqnarray}
\eta &=& \left[ {2 \hbar \over M \omega _0} \left ( {{\bf {\hat k}} \cdot \nabla \Omega ({\bf R}) \over \nu} \right )^2 - 1 \right]^{1/2} \, .
\end{eqnarray}
This means, the ratio $\eta$ is constant when the gradient of the electric field $\nabla _{\bf R} \Omega$ in the direction ${\bf {\hat k}}$ of the quantized motion of the atom experiences the same time dependence as the phonon frequency $\nu$. Now  the differential equations (\ref{dotcj2}) can be solved analytically. Given the same initial conditions as in Section \ref{initial}, i.e.~$m(0)=m_0$ and $Y_3 (0)=Y_4(0)=0$, this yields
\begin{eqnarray} \label{shrink4}
m(t) &=& m_0 + m_0 \, { \left ( 1 + \eta ^2 \right )^2 \over 2 \eta ^2} \,  \sinh ^2 \left [ \int _0 ^ t {\rm d} \tau \, \lambda (\tau) \right ] \, .
\end{eqnarray}
In case of time-independent system parameters, this equation simplifies to Eq.~(\ref{fullsolution2}). More importantly, the time derivative $\dot m$ of Eq.~(\ref{shrink4}) is always positive, independent of how $\lambda$ and $\nu$ change in time. 

\begin{figure}[t]
\begin{minipage}{\columnwidth}
{\includegraphics[]{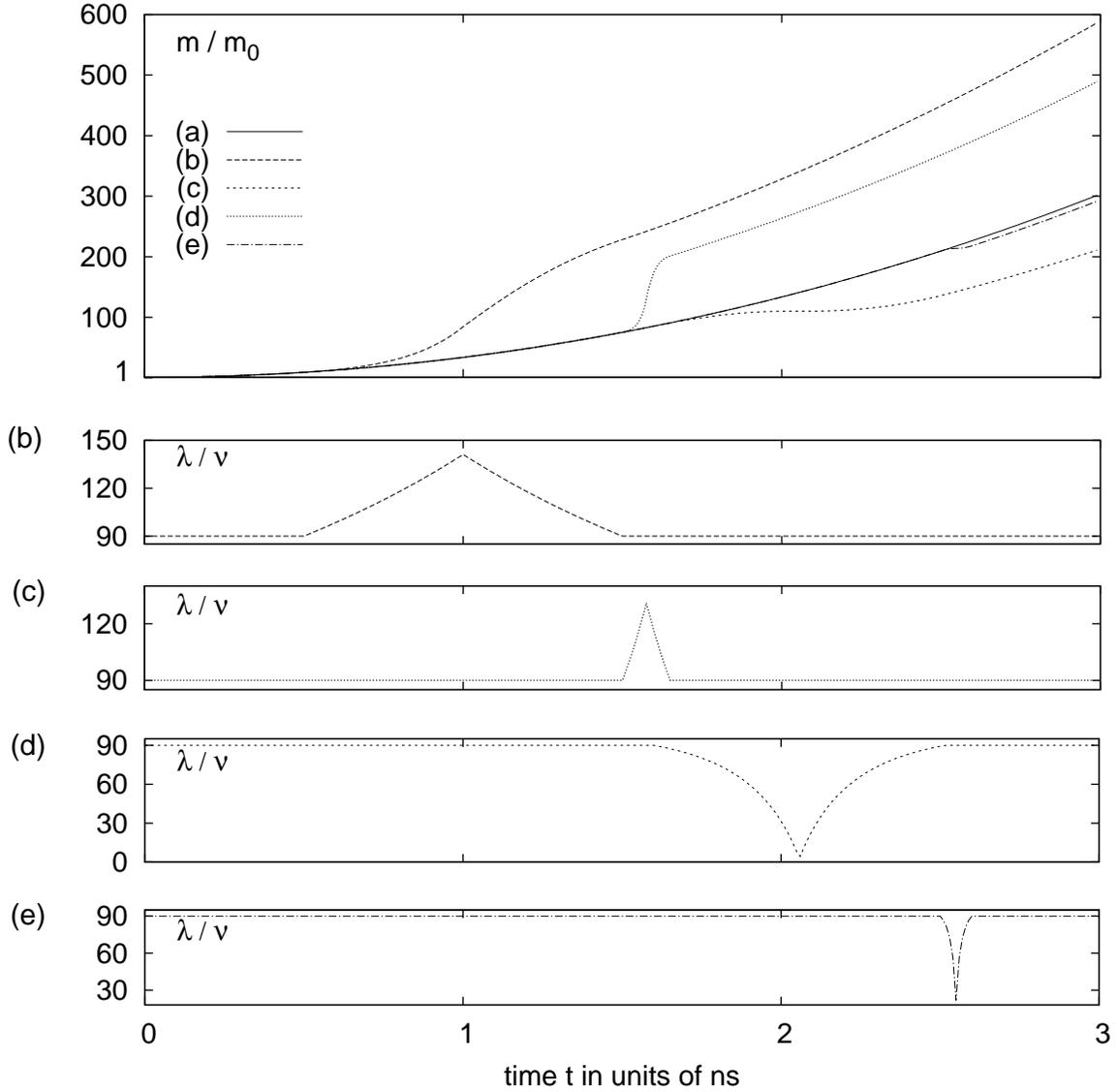}}
\vspace*{-0.5cm} \caption{Time evolution of the mean phonon number $m$ in units of the initial phonon number $m_0$ for different sudden variations of $\lambda$ and the phonon frequency $\nu$ obtained from a numerical solution of the differential equations (\ref{dotcj2}). In (a) we have $\lambda = 90 \, \rm{MHz}$, $\nu = 1 \, \rm{MHz}$ and both remain constant in time. In (b)--(d) however, $\lambda$ and $\nu$ change suddenly in time as shown with $\lambda(t) = 90 \, \nu(t)$. Here $\eta(t)$ is as in Eq.~(\ref{nu(t)}) with (b) $\kappa = 0.9 \, \rm{GHz}$ and $\Delta t = 1 \, \rm{ns}$, (c) $\kappa = 5 \, \rm{GHz}$ and $\Delta t = 0.15 \, \rm{ns}$, (d) $\kappa = -0.5 \, \rm{GHz}$ and $\Delta t = 1 \, \rm{ns}$, and (e) $\kappa = -3 \, \rm{GHz}$ and $\Delta t = 0.1 \, \rm{ns}$.} \label{figure}
\end{minipage}
\end{figure}

It is even possible to show that rapid changes of the system parameters can significantly enhance the quantum optical heating process which we analyse in this paper. To do so we now consider sudden variations of $\eta$ and ask how they affect the time evolution of $\dot m$. More concretely, we assume that the time dependence of $\eta$ can be described by an equation of the form
\begin{eqnarray} \label{shrink7}
\eta(t) &=& \eta_0 \, {\rm e}^{\kappa (t - t_0)} 
\end{eqnarray}
for times $t \ge t_0$. When combining Eqs.~(\ref{lambda}), (\ref{set}) and (\ref{shrink6}), we find that
\begin{eqnarray} \label{dotm}
\dot m & = & - {\lambda \over \sqrt{5}} \, ( c_2 + c_3 ) \, .
\end{eqnarray}
Let us now have a closer look at the differential equations for $c_2$ and $c_3$. For sufficiently rapid parameter changes, i.e.~for $|\kappa | \gg \lambda$, we can safely assume that $\eta $ changes on a much shorter time scale than $1/\lambda$. This allows us to neglect the terms $2 \lambda \, c_2$ and $- 2 \lambda \, c_3$ in Eq.~(\ref{dotcj2}). Moreover we assume that either $\lambda \gg \nu$ or $\nu \gg \lambda$. Then Eq.~(\ref{dotcj2}) can be used to obtain
\begin{eqnarray}
{{\rm d} \over {\rm d}t} (c_2 + c_3) & = & \pm {{\rm d} \over {\rm d}t} \left[ \, \ln \eta(t) \, \right] (c_2 + c_3) \, ,
\end{eqnarray}
where the plus sign applies when $\lambda \gg \nu$ and the minus sign when $\nu \gg \lambda$. This equation can be solved analytically for $\eta$'s as in Eq.~(\ref{shrink7}). Doing so and substituting its solution into Eq.~(\ref{dotm}) we obtain
\begin{eqnarray} \label{a1}
\dot m (t) &=& {\lambda(t) \over \lambda(t_0)} \, \dot m (t_0) \, {\rm e}^{\pm \kappa (t - t_0)} \, .
\end{eqnarray}
From this equation we see that the derivative $\dot m$ remains always positive as long as $\dot m (t_0)$ is positive. For $\kappa >0$ and $\lambda \ll \nu$, for example, the increase of the mean phonon number $m$ slows down when the variation of the system parameters occurs. However, it is also possible that the rate $\dot m$ increases exponentially in time. This applies, for example, when $\kappa >0$ and $\lambda \gg \nu$.

The above analytical result is in good agreement with the numerical solutions of the rate equations (\ref{dotcj2}) shown in Fig.~\ref{figure}. In this figure, we consider four different examples of the scenario where $\eta$ undergoes a rapid variation within a small time interval $(t_0,t_0+\Delta t)$. In analogy to Eq.~(\ref{shrink7}) we assume 
\begin{eqnarray} \label{nu(t)}
\eta (t) &=& \left[ {\rm e}^{\kappa (t - t_0)} -1 \right] \eta_0 \theta _{t_0, t_0 + {1 \over 2} \Delta t} (t) 
+ \left[ {\rm e}^{- \kappa (t - t_0- \Delta t )} -1 \right] \eta_0 \theta _{t_0 + {1 \over 2} \Delta t, t_0 + \Delta t} (t) \nonumber \\
\end{eqnarray}
for times $t$ between $t_0$ and $t_0+\Delta t$. Here $\theta_{t_1,t_2}(t) = 1$ if $t_1 \le t < t_2  $ and $\theta _{t_1,t_2} (t) = 0$ for any other $t$. As expected, Figs.~\ref{figure}(b) and (c) show an enhanced growth of the mean phonon number $m$ compared to the case of constant system parameters which is shown in Fig.~\ref{figure}(a). In Figs.~\ref{figure}(d) and (e), the growth of the phonon number $m$ slows down. Most importantly, the calculations in this section confirm that rapidly changing system parameters can yield much larger temperature changes inside the bubble than the ones already predicted in the previous section for time-independent parameters.

\section{Background emission in the optical regime} \label{B}

For simplicity, we now consider again time-independent system parameters. In this case, $X_3 = -1$ to a very good approximation (cf.~Eq.~(\ref{X3})) and the atom remains predominantly in its ground state. However, more detailed calculations, going up to second order in $1/\omega_0$, show that a small amount of population builds up in the excited state $|1 \rangle$ throughout the heating process. We denote this population in the following by $P_1$ and notice that
\begin{eqnarray} \label{PP1}
P_1 &=& (X_3+1)/2 \, . 
\end{eqnarray}  
To calculate $P_1$ for times $t$ of the order of $1/\lambda$, we have a closer look at the differential
equation for $X_3$ in Eq.~(\ref{fullset}). Solving it via an adiabatic evolution based on the assumption of a sufficiently large spontaneous decay rate $\Gamma$, i.e.
\begin{eqnarray} \label{Gamma}
\Gamma &\gg & \nu, \, \Omega, \, \Lambda \, ,
\end{eqnarray}
and combining the result with Eq.~(\ref{PP1}) we find 
\begin{eqnarray} \label{P}
P_1 & = & {\Omega X_2 + \Lambda Z_2 \over \Gamma} \, .
\end{eqnarray}
To calculate  $X_2$ and $Z_2$ up to second order in $1/\omega_0$, we now consider the second derivatives of the fast evolving variables $X_1$, $X_2$
and the $Z$ coherences and notice that they evolve much faster than all other
second derivatives. In fact, they are of the order of $\omega_0^2$. This allows us to adiabatically
eliminate $\dot X_1$, $\dot X_2$ and the $\dot Z$'s from the time evolution of
the system. Using Eqs.~(\ref{fullset2}) and (\ref{X3}), this yields $X_2 = 0$
and $Z_2 = - 4 \nu \Lambda Y_4/\omega_0^2$ up to second order in
$1/\omega_0$. Combining these expressions with Eqs.~(\ref{set}) and (\ref{P}), one can then
show that 
\begin{eqnarray} \label{P1}
P_1 &=& {2 \nu \over \Gamma \omega_0} \, \dot m
\end{eqnarray}
to a very good approximation. The signature even of a small but non-negligible population in the excited states of the atoms is the spontaneous emission of light in the optical regime at a measurable photon rate $I$. The reason for this is that $I$
is the sum of the contributions of all the atoms involved in the proposed heating process. If we denote the relevant atom number by $N$, then $I = N \Gamma \, P_1$ which results together with Eq.~(\ref{P1}) in
\begin{eqnarray} \label{I}
I &=& {2 N \nu \over \omega_0} \, \dot m \, .
\end{eqnarray}
This means, the intensity of the emitted light $I$ via the 0--1 transition of the nobel gas atom is proportional to the phonon frequency $\nu$ and to the heating rate $\dot m$ but does not depend on the mean phonon number $m$.

\begin{figure}
\begin{center}
{\includegraphics[]{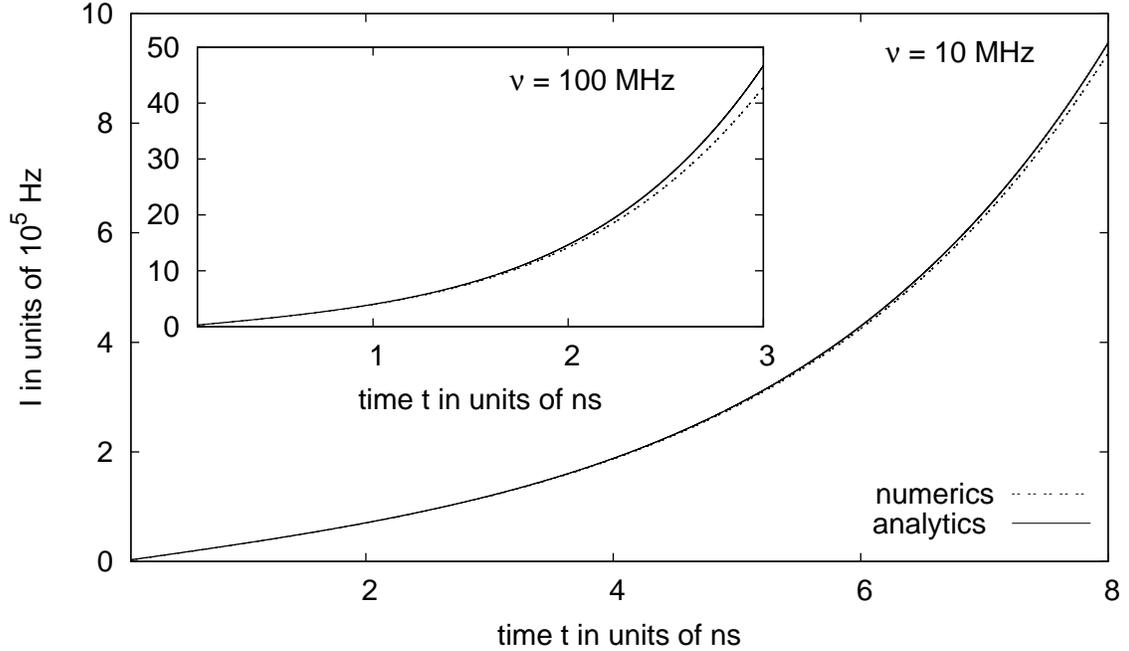}} 
\vspace*{-0.5cm} \caption{Intensity of the emitted light per atom, $I/N$, as a function of time for an initial phonon number of $m_0 = 10^7$, $\nu = 10 \,$MHz and $\nu = 100 \,$MHz, and $\omega_0$, $\Gamma$, $\Lambda$, and $\Omega$ as in Fig.~\ref{fig3}. Good agreement is found between the numerical solution of the full rate equations (\ref{fullset}) and (\ref{fullset2}) and the analytical expression in Eq.~(\ref{I}).} \label{fig6}
\end{center}
\end{figure}

Fig.~\ref{fig6} shows a very good agreement of this analytical result with the photon emission rate $I$ obtained from a numerical solution of the rate equations (\ref{fullset}) and (\ref{fullset2}). Moreover, the figure shows that $I$ increases rapidly during the collapse phase of the bubble. We see an approximately exponential growth of $I$ as a function of time. Unfortunately, the actual emission of a photon might transfer an atom into a state $|0' \rangle$ which differs from the state $|0 \rangle$ of Fig.~\ref{fig2} for example by the spin quantum number. Eventually, this results in a complete depletion of the electronic states $|0 \rangle$ and $|1 \rangle$. Then the coherences in Eq.~(\ref{coh})
become zero, thereby ending the heating process and the emission of light from the corresponding nobel gas transition. 

Notice that the light emission described in this subsection is not a description of the strong light flash at the end of the bubble collapse phase. This light flash and its spectrum are dominated by thermal radiation which can be attributed to very high temperatures inside the bubble. Here we are interested in a mechanism which might contribute substantially to the creation of these very high temperatures. The photons which we describe in this section accompany the proposed quantum optical mechanism as background radiation. Their detection would constitute a first indirect proof of our model. The background photons should be present prior to the light flash during the bubble collapse phase. They are expected to occur on a nanosecond time scale, are highly non-thermal, and should result in sharp emission lines at a frequency which is typical for the noble gas atoms inside the bubble.

For the experimental parameters in Fig.~\ref{fig6} one can easily estimate that a single atom inside the bubble emits between $10^{-4}$ and $10^{-3}$ photons per nanosecond. This means, a single atom emits on average between $10^{-4}$ and $10^{-3}$ photons prior to the picosecond light flash which ends the bubble collapse phase. The intensity of the emitted background photons depends strongly on the total number of atoms $N$ involved in the heating mechanism. Although we expect this contribution to remain small compared to the $10^5$--$10^7$ photons emitted in the optical regime during the picosecond light flash \cite{Gompf,Barber4}, it should be feasible to detect them. 

\section{Possible optical enhancement of the energy concentration} \label{poss}

One way to enhance the energy concentration in sonoluminescence experiments is to create an auxiliary highly inhomogeneous electric field inside the bubble thereby increasing the coupling between the quantised motion and the electronic  states of the atoms. However, the atoms experience a very strong trapping potential, especially during the collapse phase in single-bubble sonoluminescence experiments, as one can see for example from the robustness of the spherical symmetry of the bubble during each cycle \cite{Brenner,Suslick5}. An analogy between ion trap and sonoluminescence experiments therefore suggests another, potentially more efficient approach: namely the application of a blue-detuned laser field which should excite the 0--1 transition of the noble gas atoms (or of another atomic species with sharp optical transition lines which have already been observed experimentally \cite{Suslick3}) inside the bubble. 

In ion trap experiments, an ion experiences a strong trapping potential such that its motion becomes quantised \cite{Wineland,Eschner}. Moreover, a laser field is applied and creates a coupling between the quantised motion and the electronic states of the ion. In case of a red-detuned laser, transitions where the ion becomes excited and loses a phonon are far more likely than transitions where the atom becomes excited and gains a phonon. Followed by the return of the ion to its ground state via the spontaneous emission of a photon, the overall effect is a permanent reduction of the mean phonon number in the ions motion, i.e.~cooling. 

\begin{figure}[t]
\begin{minipage}{\columnwidth}
{\includegraphics[]{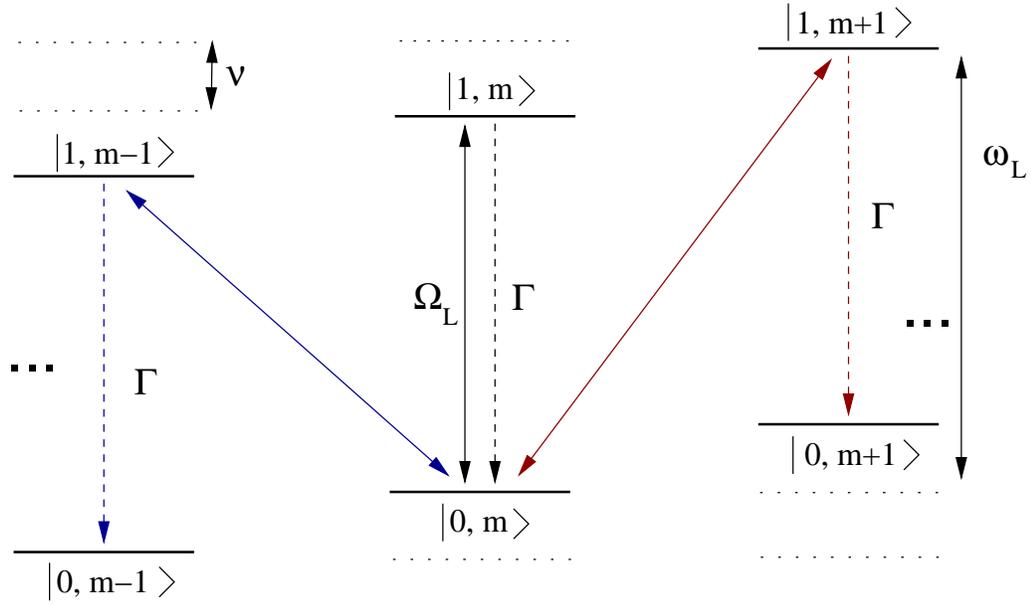}}
\caption{Possible enhancement of the energy concentration in sonolumiscence experiments via laser heating. Shown are the relevant transitions in a single atom-phonon system, when the atom is in its ground state $| 0 \rangle$, possesses exactly $m$ phonons and a laser field excites the blue sideband resonantly.} \label{fig7}
\end{minipage}
\end{figure}

The similarity of the level scheme in Fig.~\ref{fig2} and the atomic level configuration in ion trap experiments \cite{Wineland,Eschner} suggests that a laser field applied to the nobel gas atoms trapped inside the bubble in sonoluminescence experiments either enhances or suppresses the heating process, thereby testing our hypothesis of the presence of tightly trapped atoms during the bubble collapse phase. A maximum enhancement of the heating process should occur when the laser frequency equals the sum of the transition frequency $\omega_0$ and the phonon frequency $\nu$, as illustrated in Fig.~\ref{fig7}. In this case, transitions between $|0,m\rangle$ and $|1,m+1 \rangle$ are resonantly driven while transitions between $|0,m \rangle$ and $|1,m-1 \rangle$ are highly detuned. Additional laser fields should be used to avoid the premature depletion of $|0 \rangle$ and $|1 \rangle$ due to the spontaneous emission of photons into states outside the 
relevant two-level system. 

Of course, sonoluminescence experiments are very different from ion trap experiments. There are for example extreme differences in the  temperature and in the time-dependence of the trapping potential observed by each atom in its finite size environment. Ion trap experiments take place in a far more controlled environment than sonoluminescence experiments, i.e.~at temperatures close to absolute zero. Nevertheless, we expect that a laser is capable of driving the transitions in the atom-phonon systems shown in Figs.~\ref{fig2} and \ref{fig7}. Possible transitions, which can be used for this purpose, are indicated by the emission of light prior to the light flash at certain optical frequencies, as discussed in Section \ref{B}. We expect that the sonoluminescence phenomenon depends not only on the intensity of an applied laser field but also shows a very strong dependence on the laser frequency. 

\section{Conclusions} \label{conclusions}

In this paper we discuss a quantum optical heating mechanism which might contribute substantially to the sudden energy concentration in sonoluminescence experiments. Our model is based on two assumptions. First, we assume a very strong confinement of the noble gas atoms inside the bubble during the light emission phase such that their motion becomes quantised just before the maximum compression of the bubble in single-bubble sonoluminescence experiments and during rapid bubble deformations in multi-bubble sonoluminescence experiments. Secondly, our model requires the presence of a weak but highly inhomogeneous electric field. The electric field establishes a strong coupling between the quantised motion and the electronic states of each atom. Heating is due to the fact that processes which create a phonon are more likely than processes which annihilate a phonon. 

Our model does not contradict current models for the description of sonoluminescence experiments, but explains certain controversial aspects of this phenomenon. We show that the phonon energy in the bubble can easily increase by a factor of ten or more (cf.~Fig.~\ref{fig3}(c)). The relation $m \, \hbar \nu = k_{\rm B} T$ can be used to estimate the corresponding temperature increase. In fact, our model can easily predict temperatures well above $10^4\,$K inside the bubble. Rapid changes of the system parameters might further enhance the proposed quantum optical heating process. Moreover, our calculations emphasize the role of the noble gas atoms inside the bubbles. Although predominantly in their ground state, they act like catalysers and facilitate the creation of phonons at a very high rate. 

Our model is based on a quantum optical approach which is routinely used to describe the laser cooling of tightly trapped ions. To test our hypothesis of a strongly confined noble gas atoms inside the bubble during the bubble collapse phase, we propose to manipulate them with the help of an external laser field. Such a laser would create a coupling between the quantised motion and the electronic states of the atoms similar to the coupling created by a weak but highly inhomogeneous electric field. Depending on its laser frequency with respect to the relevant noble gas transition (or with respect to a sharp optical transition of another atomic species \cite{Suslick3}), it is therefore expected to either enhance or inhibit the sonoluminescence phenomenon. 

The possibility to increase the temperature inside the bubble in typical sonoluminescence experiments with the help of appropriate laser fields would be a direct proof of our model. A first indirect proof could be the observation of a background radiation prior to the light flash which accompanies the proposed quantum optical heating process. Its intensity should not depend on the temperature inside the bubble but is proportional to the heating rate $\dot m$, the phonon frequency $\nu$ and the number of nobel gas atoms involved in the heating process. The frequency of the emitted photons should correspond to typical nobel gas transitions in the optical regime.

\ack
A. B. acknowledges a James Ellis University Research Fellowship from the Royal Society and the GCHQ. This work was supported by the EU Research and Training Network EMALI and the UK Research Council EPSRC. \\


\begin{thebibliography}{10}
\bibitem{footnote}
The assumption of a relatively high spontaneous decay rate can be justified by
the presence of collective effects inside the van der Waals gas formed by the noble gas atoms and ions.
\end{thebibliography}
\end{document}